\documentclass[twocolumn,showpacs,preprintnumbers,amsmath,amssymb]{revtex4}
\usepackage[dvips]{graphicx}
\usepackage{dcolumn}
\usepackage{color}

\begin{document}

\title
{Field-induced reentrant superconductivity in thin films of 
nodal superconductor
}

\author{M. Hachiya$^{1}$, K. Aoyama$^{1,2}$, and R. Ikeda$^{1}$}

\affiliation{Department of Physics, Graduate School of Science, Kyoto University, Kyoto 606-8502, Japan$^{1}$ \\
The Hakubi Center for Advanced Research, Kyoto University, Kyoto 606-8501, Japan$^{2}$ 
}

\date{\today}

\begin{abstract} 
Previous works on nodal $d$-wave superconductors have shown that a Fulde-Ferrell-Larkin-Ovchinnikov (FFLO) like modulated superconducting (SC) state can be realized with no magnetic field when quasiparticles acquire an additional linear term in the wavevector in their dispersion. In the present work, stability of such a novel modulated SC state in an artificial film against an applied magnetic field is studied. As a reflection of the presence of the two different FFLO states, one close to zero field and the other at the high field end, in a single field v.s. temperature phase diagram of thin films, the conventional uniform SC state generally tends to appear as a reentrant ordered phase bounded by the normal phase in {\it lower} fields. 
\end{abstract}


%\keyword{}

\maketitle

\section{Introduction}

The conventional superconducting (SC) phase is characterized by a spatially uniform SC order parameter as a reflection of the Bose-Einstein condensation. In several situations, however, a SC ground state can have a spatial modulation in the SC order parameter even if quantized vortices are absent. Such a modulated SC state is regarded as one of the Fulde-Ferrell-Larkin-Ovchinnikov (FFLO) states \cite{FF,LO} and is usually stabilized by a population imbalance of up-spin and down-spin electrons in the momentum space, i.e., by a {\it uniform} splitting between the two species of Fermi surfaces provided by strong Pauli paramagnetic pair-breaking. It is believed at present through extensive studies that the high field SC phase in the heavy fermion material CeCoIn$_5$ is the above-mentioned FFLO vortex state with a one-dimensional modulation parallel to the applied magnetic field \cite{1151,
1152,Adachi}. 

It has been clarified recently in different contexts that a similar but different type of spatially modulated SC state becomes possible in nodal $d$-wave superconductors if the quasiparticle dispersion $\varepsilon_{\bf k}$ includes an additional term linear in ${\bf k}$ of the type
\begin{equation}
\varepsilon_{\bf k} = \varepsilon_{\bf k}^{(0)} + {\bf V}\cdot{\bf k}, 
\label{Rashba}
\end{equation}
where ${\bf k}$ is the wavevector, ${\bf V}$ is ${\bf k}$-independent, and $\varepsilon_{\bf k}^{(0)}$ is an even function of ${\bf k}$. Such contexts include (1) zero field cases in an external current \cite{Doh}, (2) thin films in zero field \cite{Vorontsov}, and (3) Rashba noncentrosymmetric superconductors in a magnetic field parallel to a gap node direction \cite{Kaur,Hiasa}. The vector ${\bf V}$ in eq.(1) is proportional to the strength of the external current in the case (1), the inverse of the film thickness in the case (2), and the Zeeman energy and the spin-orbit coupling in the case (3). In all of these cases, an interplay between the ${\bf k}$-dependence in the last term of eq.(1) and the corresponding one of the SC gap function favors a spatial modulation parallel to a gap-node direction in the SC state. We note that it is a FFLO-like state accompanied only by a pure {\it phase} modulation which may be realized in all the cases if the vortices are absent. 

In the present work, we focus on the above-mentioned case (2), i.e., a SC film sample with an anisotropic nodal pairing. In Ref.\cite{Vorontsov}, a tendency of formation of the modulated SC state in a SC film with $d_{x^2-y^2}$-wave pairing has been pointed out in zero field case \cite{Vorontsov}. There, a $d_{x^2-y^2}$-wave paired SC film sample is assumed to have a gap node parallel to the film plane and to be obtained by cutting the bulk sample along the plane formed by a nodal (e.g., ($1$, $1$, $0$)) and $z$ directions, and, in addition, the specular condition on the boundary surfaces is assumed. The main point in Ref.\cite{Vorontsov} is that a combination of this boundary condition and the $d$-wave pairing symmetry leads to a spatial modulation of the SC order parameter parallel to the film plane {\it and} a gap node direction. In contrast, if the boundary surface is not parallel to any gap node direction but is parallel to an anti-node (e.g., ($1$,$0$,$0$)) direction in the $d_{x^2-y^2}$-pairing case, no stable modulated SC state appears, and just the ordinary spatially uniform SC state is realized in zero field. 

Then, it is natural to imagine whether this nontrivial mechanism leading to a modulated SC state is affected by applying a uniform magnetic field. Since, at least, the Zeeman effect on the conduction electrons breaks a spin-singlet pair, the modulated state in zero field would be suppressed with increasing field. However, it is unclear whether, by increasing the field, this state gives way to the normal state or other SC states. In fact, the {\it conventional} FFLO state due to the momentum-independent population imbalance should be realized in high fields and at low temperatures. The resulting field-induced competition between the two different modulated states may lead to unusual field dependences of the phase diagram in intermediate fields. 

Hereafter, we investigate possible field v.s. temperature phase diagrams of films of unconventional superconductors in a magnetic field parallel to the film plane by assuming that the material is close to the Pauli limit so that the presence of vortices is negligible. Even in film samples with thickness of several times of the zero temperature coherence length $\xi_0$, field-induced vortices might not be negligible in general. Nevertheless, if the applied magnetic field, i.e., the straight vortex axis, is parallel to the modulation direction of the phase-modulated state occurring even in zero field, the vortex lattice pattern in the plane perpendicular to the field is unaffected by this phase-modulation, implying that conclusions on the (mean field) transitions to the normal phase and to the modulated SC states obtained in the Pauli limit, i.e., without the orbital pair-breaking, are applicable to describing real superconductors with large paramagnetic pair-breaking effect. This argument is based on previous studies on the corresponding bulk system \cite{Adachi,com1}. Of course, the spatially uniform SC phase appearing in our calculation in the Pauli limit should correspond to an ordinary vortex lattice in a thin film, which consists of the vortices running straight along the field parallel to the film plane. 

Geometry of the present system is illustrated in Fig.1. Although our focus is primarily on the nodal $d$-wave superconductors, we also discuss a simpler nodal $p$-wave pairing case in order to clarify that one of origins of the present phase-modulated state in zero field is a gap node parallel to the film plane. It is found that, for a film thickness of several times of $\xi_0$, applying a weak parallel magnetic field in the $d$-wave nodal SC case changes the modulated SC state in {\it low} fields to the normal state, while the uniform SC and the {\it conventional} FFLO states survive in higher fields further. Consequently, reentrant SC phases are expected to occur with increasing the field for a film thickness of several times of $\xi_0$. It is pointed out that such a strange field dependence of the SC phases, in part, stems from the nonmonotonous thickness dependence of the SC transition temperature in zero field \cite{Vorontsov}. 

%%%%%%%%%%%%%%%%%%%%%%%%%%%%%%%%%%%%
\begin{figure}[t]
\includegraphics[scale=0.38]{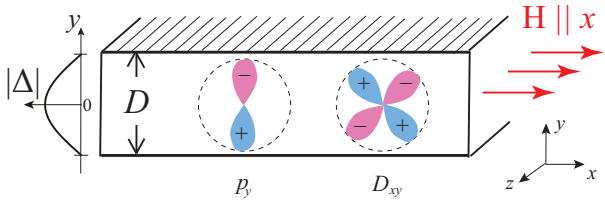}
\caption{Geometry of superconducting (SC) film samples with thickness $D$ and the $d_{x^2-y^2}$-pairing in which the gap nodes are parallel to the film plane (the $d_{x^2-y^2}$ symmetry in this setup is called hereafter as $D_{xy}$). The $p_y$-pairing case is also indicated here for comparison. The flat film surfaces are in $z$-$x$ plane. Due to the specular boundary condition on the quasiparticles under the gap nodes parallel to the film plane, the amplitude of the SC gap function $|\Delta|$ is suppressed to zero on the sample surfaces. Effects of a finite magnetic field applied in the gap-node direction within the 2D plane (the $x$-direction ) will be considered in sec.III.}
\end{figure}
%%%%%%%%%%%%%%%%%%%%%%%%%%%%%%%%%%%%%%

The present paper is organized as follows. In sec.II, theoretical formulation is explained, and thickness v.s. temperature phase diagrams of a $d$-wave film taken at low enough fields are discussed together with a $p$-wave case in zero field in sec.III. In sec.IV, field v.s. temperature phase diagrams in the $d$-wave case which are our main focus in the present work are discussed. In sec.V, the present work is summarized.

\section{Formulation}

In the present work, the boundary condition for the SC order parameter $\Delta$ on the film surface plays an important role in obtaining a spatial modulation of the SC order parameter. To explain how the boundary condition results in a modulated SC phase, let us start from describing the quasiclassical approach used in the present work. 

Since we study not only the spin-singlet $d$-wave pairing in magnetic fields but also a $p$-wave pairing case in zero field, Eilenberger equations for the quasiclassical Green's functions \cite{Eilenberger,LO1,Schopohl} represented in the Nambu space will be considered. Following Ref.\cite{Luttinger_1,VSG}, they are expressed by 
\begin{equation}
[i \varepsilon_n {\hat \tau}_3 - {\hat \Delta} - {\hat I}, \,\,\, {\hat g} ] + i {\bf v}_{\rm F}\cdot\nabla {\hat g}=0,
\end{equation}
where the bracket denotes the commutator, $\varepsilon_n$ is a fermion Matsubara frequency, and ${\hat \tau}_j$ ($j=1$, $2$, and $3$) are the particle-hole Pauli matrices. 
We largely follow Ref.\cite{VSG} regarding the notation of the gap function and the Green's functions ${\hat g}$ satisfying $({\hat g})^2 = - \pi^2$. They will be parameterized in the form 
\begin{widetext}
\begin{eqnarray}
{\hat \Delta} &=& \left( \begin{array}{cc} 0& i(\Delta({\bf r}, {\bf p}) + \sum_{j=1,2,3} {\bf \Delta}_j({\bf r},{\bf p})\cdot\sigma_j)\sigma_2  \\ i\sigma_2(\Delta^*({\bf r}, {\bf p}) + \sum_{j=1,2,3} {\bf \Delta}^*_j({\bf r},{\bf p})\cdot\sigma_j) &0\end{array} \right), \quad \nonumber \\
{\hat g} &=& \left( \begin{array}{cc} g+{\bf g}\cdot{\bf \sigma}& i(f+{\bf f}\cdot{\bf \sigma})\sigma_2 \\ i\sigma_2(f'+{\bf f}'\cdot{\bf \sigma})&-g+{\bf g}\cdot{\bf \sigma}*\end{array} \right). 
\end{eqnarray}
\end{widetext}
The matrix ${\hat I}$ expresses the Zeeman energy term. In the parallel field configuration (${\bf H} \perp {\hat y}$) of our interest in the following sections, ${\hat I}$ is diagonal, and its element gives the Zeeman energy $\mu H$ in the Nambu space. Further, ${\bf r}=(x,y,z)$ is the center-of-mass coordinate of a Cooper pair. As mentioned in sec.I \cite{com1}, the orbital pair-breaking effect inducing the vortices can be safely neglected in the present geometry (see Fig.1). Thus, the strength of the magnetic field can be identified hereafter with the dimensionless Zeeman energy 
\begin{equation}
h \equiv \frac{2}{\pi} e^\gamma \, \frac{\mu \, H}{T_{c}(0)},
\end{equation}
where $\gamma$ is the Euler constant. The numerical factor $4 e^\gamma \simeq 7.1$ in the above expression results from the conventional definition of the Maki parameter $\alpha_{\rm M}$ for the $s$-wave bulk superconductor \cite{Maki}. 

As is well known, minimization of the free energy with respect to the gap function $\Delta$ results in the so-called gap equation, which, in the spin-singlet-pairing case, connects ${\hat \Delta}$ with the scalar anomalous Green's functions $f(\varepsilon_n)$ 
and $f'(\varepsilon_n) = (f(-\varepsilon_n))^*$ in the manner 
\begin{eqnarray}
\Delta({\bf r}) \, {\rm ln}\biggl(\frac{T_{c0}}{T}\biggr) &=& T \sum_{n \geq 0} \biggl\langle \text{$\cal{Y}$}^*({\bf p}) \biggl(\frac{2 \pi \Delta({\bf r}) \text{$\cal{Y}$}({\bf p})}{|\varepsilon_n|} \nonumber \\
&-& f({\bf r}, {\bf p}; \varepsilon_n) - f^{' \, *}({\bf r}, {\bf p}; \varepsilon_n) \biggr) \biggr\rangle_{\bf p}. 
\label{GAP}
\end{eqnarray}
Here, $T_{c0}$ is the SC transition temperature in zero field, and $\langle \,\,\,\,\rangle_{\bf p}$ is the angle average over the Fermi surface. Further, ${\cal Y}({\bf p})$ denotes the normalized pairing function satisfying $\langle |\text{$\cal{Y}$}({\bf p})|^2 \rangle_{\bf p}=1$. It appears through the factorization $\Delta({\bf r}, {\bf p}) = \Delta({\bf r}) {\cal Y}({\bf p})$ and, in the case with gap nodes parallel to the ${\hat x}$-direction, is proportional to ${\hat p}_y$.

The system geometry is reflected in the boundary condition on $f({\bf r}, {\bf p}; \varepsilon)$ on the outer surface. 
As in Ref.\cite{Vorontsov}, the specular condition on a flat boundary surface in $z$-$x$ plane will be used by assuming a surface roughness to be negligible (see the geometry in Fig.1). Then, we have 
\begin{equation}
f({\bf r}; {\bf p}; \varepsilon_n) 
= f({\bf r}; {\overline {\bf p}}; \varepsilon_n)
\label{BCoff}
\end{equation}
{\it on the surface}, where ${\overline {\bf p}}={\bf p} - 2 {\hat {\bf n}}({\bf p}\cdot{\hat {\bf n}})$, and ${\hat {\bf n}}$ is the unit vector normal to the surface, which is in the $y$-direction in Fig.1. It means that, if 
\begin{equation}
{\cal Y}({\overline {\bf p}}) = - {\cal Y}({\bf p}), 
\label{BC}
\end{equation}
the SC order parameter $\Delta({\bf r})$ and $f$ vanish on the surface.
%where $\Delta({\bf r}, {\bf p})={\cal Y}({\bf p}) \Delta({\bf r})$. 
Among pairing states satisfying eq.(\ref{BC}), we will focus below on the following ones 
\begin{eqnarray}
{\cal Y}({\bf p}) &=& \sqrt{2} {\hat p}_x {\hat p}_y, \nonumber \\
{\cal Y}({\bf p}) &=& \sqrt{2}{\hat p}_y, 
\label{exam}
\end{eqnarray}
both of which, in $x$-$y$ plane, have a gap node in the parallel direction to the boundary surface. The former case corresponds to a film sample prepared by cutting a $d_{x^2-y^2}$-paired superconductor along, for instance, the ($1$,$1$,$0$) direction. Hereafter, this situation with the $d_{x^2-y^2}$-pairing symmetry will be called as $D_{xy}$-pairing case. 

In addition, as a typical ${\bf r}$-dependence of $\Delta({\bf r})$ vanishing on $y=\pm D/2$ (see Fig.1), we choose the form 
\begin{equation}
\Delta({\bf r}) 
= \Delta(x) \, \sqrt{2} {\rm cos}(\pi y/D). 
\end{equation}
For simplicity, the cylindrical Fermi surface with the symmetry axis parallel to ${\hat z}$ will be used throughout this paper. 
%Below, we also consider the spin-triplet $p_y$-pairing case, the latter of eq.(%\ref{exam}), as another simpler model leading to a modulated SC state in zero 
%field. 

As will be stressed in sec.III, a continuous normal to superconducting (SC) transition point can be found directly from the O($|\Delta|^2$) term of the free energy, or equivalently, from eq.(\ref{GAP}) linearized with respect to $\Delta$. To obtain a transition occurring deep in the SC phase such as an onset of a FFLO modulation, the following quasiclassical representation \cite{VSG} of the SC contribution $\Delta F_{\rm sc}$ to the Luttinger-Ward free energy functional 
\begin{equation}
\Delta F_{\rm sc} = \int dxdy \frac{T}{2} \int_0^1 d\lambda N(0) \sum_{\varepsilon_n} \biggl\langle {\rm Tr} {\hat \Delta} \biggl({\hat g}_\lambda - \frac{1}{2} {\hat g} \biggr) \biggr\rangle_{\bf p} 
\label{free energy}
\end{equation}
is more useful rather than the gap equation (\ref{GAP}), where $N(0)$ is the density of states (per spin) at the Fermi level in the normal state, and the auxiliary Green's function ${\hat g}_\lambda$ is the solution of the Eilenberger equation with ${\hat \Delta}$ replaced by $\lambda {\hat \Delta}$. The transition from the uniform SC state to a phase-modulated Fulde-Ferrell (FF) SC state 

\begin{equation}
\Delta_{\rm FF}(x) = \exp(iq_x x) |\Delta(0)|
\end{equation}

or to an amplitude-modulated Larkin-Ovchinnikov (LO) SC state 

\begin{equation}
\Delta_{\rm LO}(x) = \sqrt{2} {\rm cos}(q_x x) |\Delta(0)|
\end{equation}

is signaled by emergence of a finite equilibrium value of the modulation wave number $q_x$. 

\section{Thickness v.s. Temperature Phase Diagrams}
%In the remainder of this section, we will primarily discuss the zero field case% \cite{Vorontsov}. 

In this section, we will discuss the zero field case. By comparing thickness-temperature phase diagrams for the spin-triplet $p_y$-pairing and the spin-singlet $D_{xy}$-pairing \cite{Vorontsov} cases to each other, we stress that the presence of a gap node parallel to the film plane induces the phase-modulated state at lower temperatures and that a strange thickness dependence of the SC transition temperature is commonly seen in those SC films. 

To discuss the character of the normal to SC phase transition in the mean field (MF) approximation which is signaled by vanishing of the SC energy gap, the form of the Ginzburg-Landau (GL) free energy which can be obtained from the gap equation (\ref{GAP}) will be explained. By introducing the Fourier-transformation $\Delta(x)=L_x^{-1/2} \sum_{q_x} \Delta_{q_x} \exp(iq_x x)$ to incorporate possibilities of a spatially varying mean field solution of $\Delta(x)$, the quadratic (O($|\Delta|^2$)-) term of the GL free energy is recovered 
in the form 
\begin{widetext}
\begin{eqnarray}
F_{H=0} = N(0) \sum_{q_x} \biggl[ \,{\rm ln}\biggl(\frac{T}{T_{c0}} \biggr) + \int_0^\infty d\rho \frac{2 \pi T}{{\rm sinh}(2 \pi T \rho)} \biggl\langle |{\cal Y}({\bf p})|^2  \biggl( 1 
- {\rm cos}(\rho v_F {\hat p}_x q_x) {\rm cos}\biggl(\rho \pi v_F \frac{{\hat p}_y}{D} \biggr) \biggr) \biggr\rangle_{\phi_{\bf p}} \, \biggr] |\Delta_{q_x}|^2
\label{GL2}
\end{eqnarray}
\end{widetext}
from the O($\Delta$) term of eq.(\ref{GAP}), where $\langle \,\,\, \rangle_{\phi_p}$ denotes the average over $\phi_{\bf p}$ with $p_x + {\rm i}p_y = p_{\rm F} \exp({\rm i}\phi_{\bf p})$ for the present cylindrical Fermi surface. Here, by first assuming the transition to be continuous, let us consider the situation under which the coefficient of $|\Delta_{q_x}|^2$ in eq.(\ref{GL2}) vanishes. In the cases of our interest, ${\cal Y}({\bf p}) \propto p_y$, i.e., when we have gap nodes in the ${\hat x}$-direction, the sign factor ${\rm cos}(\rho \pi v_F {\hat p}_y/D)$ in eq.(\ref{GL2}) may contribute to the negative sign for a broad range of the $\rho$-value. Then, inevitably, a state with a finite $q_x$, i.e., a state with a modulation parallel to a gap node is favored in order for ${\rm cos}(\rho v_F {\hat p}_x q_x)$ to be also accompanied by the negative sign so that the negative sign of the coefficient of the $|\Delta_{q_x}|^2$ term, i.e., the SC ordering, may become possible. Like the mechanism leading to the conventional FFLO state induced by the Zeeman field, the dimensionless film thickness normalized by the zero temperature coherence length $\xi_0= v_{\rm F}/(2 \pi T_c)$, i.e., 
\begin{equation}
d^{-1} = \frac{v_{\rm F}}{2 T_c D} = \pi \frac{\xi_0}{D}
\label{dimlessd}
\end{equation} 
plays the roles of the Zeeman energy in the familiar field-induced FFLO case in the present case. 

%%%%%%%%%%%%%%%%%%%%%%%%%%%%%%%%%%%%
\begin{figure}[t]
\includegraphics[scale=0.45]{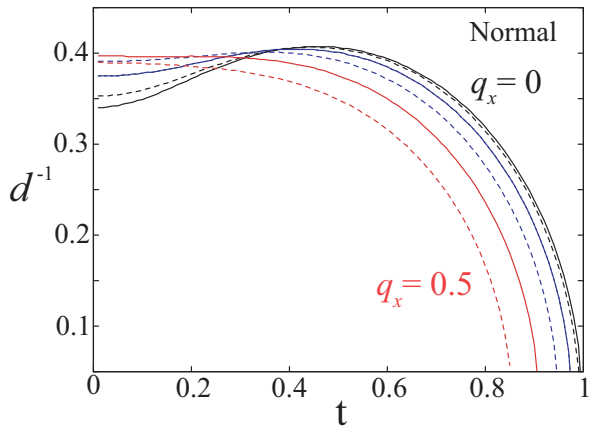}
%\scalebox{0.3}[0.3]{\includegraphics{Fig2.eps}}
\caption{Thickness dependences of the SC transition temperature, $T_c(d^{-1})$, of superconducting films with the pairing symmetry ${\cal Y}({\bf p})=\sqrt{2} {\hat p}_y$ in zero field which have been obtained by assuming various $q_x$-values, where $t=T/T_c(0)$. The envelope of those $T_c(d^{-1})$-curves is nothing but the resulting mean field SC transition curve between a SC phase and the normal (N) phase and is described as a solid curve in Fig.3 (a). At lower temperatures where the boundary condition (a finite $D$) is more effective, the $T_c(d^{-1})$-curve with finite $q_x$ tends to be realized. The SC transition is always of second order (see Fig.3).}
\end{figure}
%%%%%%%%%%%%%%%%%%%%%%%%%%%%%%%%%%%%%%
Figure 2 shows dependences of the normal to SC phase boundary on {\it assumed} $q_x$-values in the $d^{-1}$ v.s. temperature phase diagram in the $p_y$-pairing case with zero field. 
The actual mean field SC transition line is defined as the envelope of each curve giving the highest $T_c$ and $d^{-1}$ values among those curves. Its result is presented in Fig.3 as the solid black curve, which implies that, at lower temperatures and/or for larger $d^{-1}$, a SC phase with a nonzero $q_x$ leads to a higher transition line and thus, is favored as the ordered state. The resulting stability region of a modulated FF SC state has been determined by use of eq.(\ref{free energy}). We note that, as pointed out in Ref.\cite{Vorontsov}, the present phase-modulated FF state is always more stable than the LO state. This phase-modulated state corresponds to the helical state in the context of the noncentrosymmetric superconductors with {\it no} field-induced vortices accompanied \cite{Kaur,Aoyama2}. 

%%%%%%%%%%%%%%%%%%%%%%%%%%%%%%%%%%%%
\begin{figure}[t]
\includegraphics[scale=0.5]{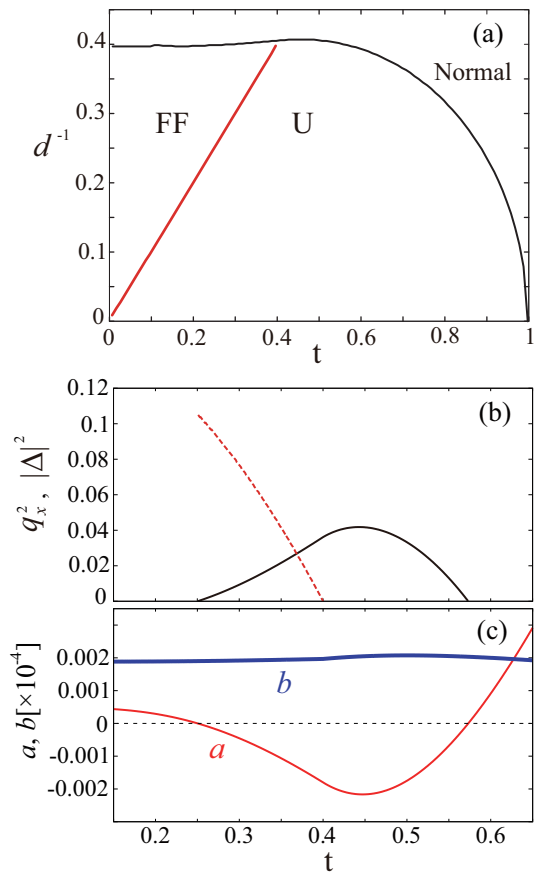}
%\scalebox{0.3}[0.3]{\includegraphics{Fig3.eps}}
\caption{(a) Resulting thickness dependence of the SC transition temperature, $T_c(d^{-1})$ (solid black curve), for the $p_y$-pairing obtained from the data in Fig.2. Although the $T_c$ v.s. $d^{-1}$ curve in $t < 0.4$ becomes nonmonotonic upon cooling (see the text), the transition on $T_c(d^{-1})$ is of second order. The SC state at lower temperatures and at smaller values of the film thickness is the phase-modulated Fulde-Ferrell (FF) state which is separated from the uniform SC (U) state via another second order transition (solid red) curve decreasing upon cooling. (b) Temperature dependences of the squared order parameters $|\Delta|^2$ and $q_x^2$ at $d^{-1}=0.399$, and (c) the corresponding coefficient $a$ of the GL quadratic term [thin (red) curve] and the corresponding one $b$ of the quartic term [thick (blue) curve] (see Appendix on their definition). Here, $\Delta$ and $q_x$ are normalized by $T_c(d^{-1}=0)$ and $\xi_0^{-1}$, respectively, and hence, are dimensionless. The resulting SC and FF transitions are of second order in character, as signaled by the linear vanishings of $|\Delta(t)|^2$ and of $(q_x(t))^2$, respectively.}
\end{figure}
%%%%%%%%%%%%%%%%%%%%%%%%%%%%%%%%%%%%%%
                                                                               
In addition, to clarify the character of this SC transition, the sign of the quartic (O($|\Delta|^4$)) term in the GL free energy needs to be examined. In fact, in the case with a {\it conventional} FFLO state induced by the ${\bf p}$-{\it independent} Zeeman energy, the overall coefficient $b$ of the quartic term at lower temperatures is negative near the line on which the quadratic term changes its sign so that the corresponding SC transition is of first order \cite{Adachi}. However, in the present cases where the counterpart of the Zeeman energy, i.e., $v_{\rm F} {\hat p}_y/D$, is linear in ${\bf p}$, it is found that the coefficient $b$ of the quartic term remains positive even at lower temperatures (see Fig.3 (c)) and thus that the mean field SC transition is truly of second order. Again, this result is compatible with the closely related result on the low temperature $H_{c2}(T)$-transition in the Rashba noncentrosymmetric superconductors under a field parallel to the basal plane \cite{Hiasa} (see Introduction). On the other hand, in the case of the present Fig.2, the resulting SC transition curve defined by a sign change of the quadratic term shows a {\it nonmonotonous} temperature dependence. This feature is in contrast to that in the {\it conventional} case with the ${\bf p}$-independent paramagnetic pair-breaking, where such a nonmonotonous temperature dependence of the curve on which the quadratic term changes its sign indicates that the mean field SC transition discontinuously occurs on a different curve \cite{Adachi}. 
%Such a nonmonotonic behavior is also seen in the SC energy gap for small $d$-va%lues and, as discussed later, significantly affects field dependences of the SC% phases of thinner films. 

%%%%%%%%%%%%%%%%%%%%%%%%%%%%%%%%%%%%
\begin{figure}[t]
\includegraphics[scale=0.5]{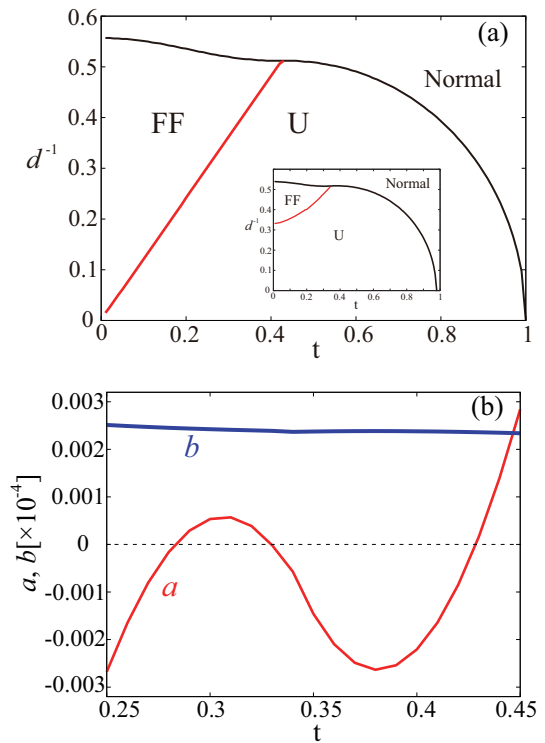}
%\scalebox{0.3}[0.3]{\includegraphics{Fig4.eps}}
\caption{(a) Thickness dependence of the SC transition temperature, $T_c(d^{-1})$, in the $D_{xy}$-pairing case. This phase diagram in $h=0$ quantitatively agrees with that in Ref.\cite{Vorontsov}. The inset shows the corresponding result in the small field $h=0.32$. The nonmonotonic behavior of $T_c(d^{-1})$ curve at low temperatures is less remarkable, possibly because of the additional gap nodes perpendicular to the film plane, but nevertheless visible close to $t=0.35$. (b) Temperature dependences of the GL coefficients $a$ [thin (red) curve] and $b$ [thick (blue) curve] at $d^{-1}=0.517$ in the figure (a). The positive $b$-values in the changes of the sign of $a$ on cooling implies that the SC transition in the figure (a) is of second order. }
\end{figure}
%%%%%%%%%%%%%%%%%%%%%%%%%%%%%%%%%%%%%%

Figure 4 (a) shows the $d^{-1}$ v.s. temperature phase diagram of the $D_{xy}$-pairing film in zero field, which coincides with the result in Ref.\cite{Vorontsov}. 
%Here $D_{xy}$-pairing film indicates a film system of a $d_{x^2-y^2}$-paired su%perconductor with a nodal direction parallel to the film plane. 
Just like that of the $p_y$-pairing in Fig.3, the uniform SC phase tends to be destabilized due to formation of the FF state with decreasing the film 
thickness. 
In the $D_{xy}$-pairing case, there are also the gap nodes perpendicular to the film plane which may partially cover the formation of the FF phase modulation stemming from the gap nodes parallel to the plane. 
In fact, although a nonmonotonic behavior of the second order transition curve $T_c(d^{-1})$ is present even in the $D_{xy}$-pairing case (see Fig.4 (b)), it is less remarkable compared with that in the $p_y$-pairing case (Fig.3). Nevertheless, as the inset of Fig.4 (a) shows, this low temperature behavior of the $T_c(d^{-1})$ curve is robust against a weak magnetic field. In sec.III, it will be shown that this nearly flat but nonmonotonic feature close to $t=0.35$ of $T_c(d^{-1})$-curve results in a reentry of the FF and normal phases in films with thickness of about $6 \xi_0$. In contrast, in sufficiently thick films with large $d$, the FF phase is destabilized by such a weak magnetic field (see the inset of Fig.4). 

For comparison, we will comment on the $d_{x^2-y^2}$-pairing case in which an anti-nodal direction is parallel to the film plane. In this case, the boundary condition (\ref{BC}) is not satisfied, and the uniform superconductivity rather than the FF state is always favored. Thus, within the approximation used here, disappearance of superconductivity due to the size effect does not occur in this case. 

%We have verified that the corresponding calculation in the case with $w_{\bf p}% = \sqrt{2} {\hat p}_x {\hat p}_y$ precisely reproduces the phase diagram in Re%f.\cite{Vorontsov}. 

%\bibitem{com} de, fluctuation effect wo iretemo, kono nonmonotonic behavior ga %kaerareru konkyo ha nai. wo ireyo. 

\section{Field v.s. Temperature Phase Diagrams} 

Next, let us examine how the phase diagram of the $D_{xy}$-pairing case is changed by applying a uniform magnetic field. For this purpose, the orbital pair-breaking effect of the magnetic field will be neglected. This assumption is reasonable for describing thin films with thickness of several times of $\xi_0$. Further, as argued in sec.I, it is at least qualitatively valid even with vortices as far as ${\bf H} \parallel {\hat x}$ \cite{com1}. Then, a nonvanishing magnetic field appears only through the Zeeman energy $\mu H$, and its effect can be incorporated simply by replacing ${\rm cos}(\rho \pi v_F {\hat p}_y/D)$ with ${\rm cos}(\rho \pi v_F {\hat p}_y/D) \, {\rm cos}(\rho 2 \mu H)$ in eq.(\ref{GL2}) (see \cite{Adachi} and Appendix for details). Then, the two cosine-factors compete in their signs in the $\rho$-integral in eq.(\ref{GL2}). Physically, this implies that the applied magnetic field frustrates and thus, weakens the size effect enhanced by the gap nodes. Since, on the other hand, superconductivity in the present systems is suppressed with decreasing the film thickness, the applied magnetic field competing with the size effect may enhance superconductivity. Below, it will be explained that, in thin SC films with gap nodes parallel to the film plane and with thickness of the order of several times of $\xi_0$, rich behaviors reflecting the competition between the magnetic field and the gap node-induced size effect can occur in the field v.s. temperature phase diagram such as coexistence of different kinds of spatially-modulated FFLO states and a field-induced reentry of the spatially-uniform superconductivity. 

%%%%%%%%%%%%%%%%%%%%%%%%%%%%%%%%%%%%
\begin{figure}[t]
%\scalebox{0.25}[0.25]{\includegraphics{Fig5a.eps}}
%\scalebox{0.25}[0.25]{\includegraphics{Fig5b.eps}}
%\scalebox{0.25}[0.25]{\includegraphics{Fig5c.eps}}
%\scalebox{0.25}[0.25]{\includegraphics{Fig5d.eps}}
%\scalebox{0.25}[0.25]{\includegraphics{Fig5e.eps}}
%\scalebox{0.25}[0.25]{\includegraphics{Fig5f.eps}}
\includegraphics[width=\columnwidth]{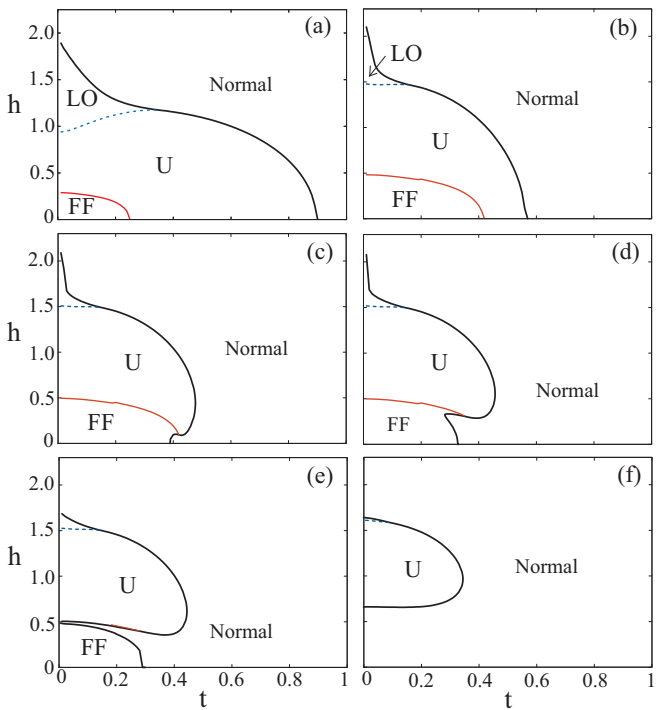}
\caption{Thickness dependence of $h$ v.s. $t$ phase diagram in the $D_{xy}$-pairing case. The used value of $d$ (the normalized thickness defined in eq.(\ref{dimlessd})) are 3.33 in (a), 2.0 in (b), 1.95 in (c), 1.93 in (d), 1.92 in (e), and 1.79 in (f), respectively. The symbol "LO" indicates the field-induced LO phase. All transitions indicated in the figures are of second order (see Fig.6). Field dependences of the order parameters on the dotted vertical lines in the figures (d) and (e) are shown in Fig.6. }
\end{figure}
%%%%%%%%%%%%%%%%%%%%%%%%%%%%%%%%%%%%%%

Thickness dependence of the $h$ v.s. $t$ phase diagrams we have obtained is presented in Fig.5. Although the obtained transition curves in the figures have been determined from the quasi-classical formulation in sec.II with eq.(\ref{free energy}), all the SC transitions appearing in the present phase diagram can be alternatively obtained from the GL free energy shown in Appendix because they are of second order in character. As seen in Fig.3 of Ref.\cite{Vorontsov} (see also Fig.4 in the present work), as far as the same boundary condition is used, the FF phase of SC films with a gap node parallel to the film plane under zero field inevitably appears even in sufficiently thick films with $d \gg 1$ at low enough temperatures \cite{Vorontsov,com2}. Reflecting this fact, Fig.5(a) includes this FF phase at low enough temperatures if the applied field is sufficiently low. With decreasing the film thickness, however, the temperature regions of the uniform and LO SC phases are narrower, while the FF phase grows and begins to occupy a broader field and temperature region (see Fig.5(b)). In particular, it is notable in Fig.5(b) that, as the FF phase grows, the field-induced LO phase shrinks and, together with the normal to SC transition curve, is pushed up to higher fields. This competition between these two modulated SC phases seems to result in extension of the intermediate {\it spatially-uniform} SC phase. For instance, the point ($t=0.2$, $h=1.4$) in Fig.5(b) is included in the uniform SC phase, although the same point in Fig.5(a) is in the normal phase, suggestive of an {\it extension} of the uniform SC phase in thinner films. More interestingly, 
for further thinner films with the thickness $D < 6.13 \xi_0$, the normal phase begins to intervene between the FF and the uniform SC phases at low temperatures to push the FF phase down to lower fields with decreasing the thickness (see Fig.5(c) and (d)). It is a remarkable feature that the normal phase begins to enter along the boundary between the FF and uniform phases with decreasing the film thickness so that the uniform SC phase at {\it higher} fields can remain stable against the normal state (see Fig.5(e) and (f)). The above-mentioned {\it competition} between the size effect and the finite magnetic field and the resulting reentrant survival of the {\it spatially-uniform} SC phase are nontrivial and main results in the present work. 

As already mentioned, all the transitions appearing in these phase diagrams are of second order. In Fig.6, field dependences of the amplitude of the SC order parameter $\Delta$ and the FFLO order parameter $q_x$ taken on the dotted vertical lines in Fig.5(d) and (e) are shown. For instance, in Fig.6 (a), the linear vanishing of $|\Delta|^2$ at three $h$-values and that of $q_x^2$ in the vicinity of $h=0.39$ imply the second order character of these transitions. The $q_x^2$ curve has its physical meaning when $|\Delta|$ is nonvanishing. For this reason, $q_x$ data are not shown in Fig.6 in the field regions of the normal state. 

%%%%%%%%%%%%%%%%%%%%%%%%%%%%%%%%%%%%
\begin{figure}[t]
\includegraphics[scale=0.5]{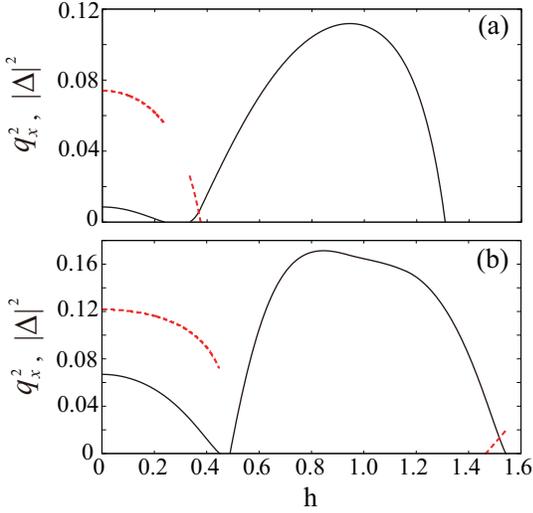}
\caption{Data of the {\it squared} order parameters $|\Delta|^2$ (solid curves) and $q_x^2$ (dashed ones) taken on the dotted vertical lines in Figs.5 (d) and (e) are presented in the figures (a) and (b), respectively. The linear vanishings of $|\Delta|^2$ and of $q_x^2$ under a finite $|\Delta|$-value ensure that these transitions are of second order in character.}
%\scalebox{0.3}[0.3]{\includegraphics{Fig6.eps}}
%\caption{Field dependences of the amplitude of the SC order parameter $|{\tilde% \Delta}|$ and the FFLO order parameter $|q_x|$ in Fig.5(d) at $t=0.3$. }
\end{figure}
%%%%%%%%%%%%%%%%%%%%%%%%%%%%%%%%%%%%%%

In contrast to the $D_{xy}$-pairing case, the FF phase does not occur if the film plane is parallel to an anti-nodal direction of the $d_{x^2-y^2}$-pairing function. Then, the resulting field v.s. temperature SC phase diagram consists only of the uniform and LO SC phases with no reentry of a low field normal phase. 

\section{Conclusion}

In this paper, we have studied the phase structure of a SC film with an unconventional nodal pairing and, in particular, have focused on its change due to an applied magnetic field parallel to the film plane. It has been found that, when the pairing state has a gap node parallel to the film plane, and consequently, the SC energy gap or the SC order parameter is deformed along the film's surface normal, a kind of Fulde-Ferrell (FF) SC phase with a spatial modulation parallel to the film plane of the {\it phase} of the SC order parameter is inevitably induced upon cooling in zero field. When a magnetic field parallel to the gap-node direction is applied, and the system is close to the Pauli limit, this unconventional FF state and the conventional Larkin-Ovchinnikov (LO) state in high fields coexist in the same field v.s. temperature phase diagram. With decreasing the film thickness, these modulated SC states seem to repel each other in the phase diagram while keeping the uniform SC phase intervening between them intact. Consequently, the FF phase in lower fields is replaced with the normal phase in thinner films so that a field v.s. temperature phase diagram with a reentrant uniform SC phase results in. 

As mentioned in sec.I, vortices can be neglected in discussing thermodynamics and the static phase diagram of superconductors close to the Pauli limit. Since, as seen in Fig.5, the main point in the present work is the reentry of the normal phase in {\it lower} fields in thinner films, inclusion of the field-induced vortices in the present theory is expected not to lead to essential changes of our main results. 

An intriguing point is an apparent competition between the Fulde-Ferrell (FF) state in lower fields and the conventional Larkin-Ovchinnikov (LO) state at the high field end of the SC phase. In fact, the FF phase arises from the anisotropic response to an external field in the momentum space (see eq.(\ref{Rashba})) and thus, is incompatible with the LO state to be realized even in the isotropic system. Therefore, it may be expected that the field-induced reentry of the uniform SC phase will also occur in other systems with a zero field FFLO phase, e.g., a case with a FFLO state induced by a multi-band effect or with a FFLO state induced by an electric current \cite{Doh} . These issues will be considered elsewhere. 

This work is supported by Grant-in-Aid for Scientific Research [No. 25400368] from MEXT, Japan.

\section{Appendix}
\subsection{Derivation of the GL quadratic term}
The Ginzburg-Landau quadratic term essentially corresponds to the linearized gap equation, so that it can be obtained by expanding eq.(5) with respect to $\Delta$ and picking up the leading order contributions. In this appendix, we sketch an alternative derivation \cite{Adachi} of the GL free energy functional based on the Gor'kov formalism which is equivalent to the $\Delta$-expansion of eq.(5). The GL quadratic term is formally given by
\begin{eqnarray}\label{eq:GL}
a |\Delta(0)|^2 &=& \frac{1}{N(0)} \int \frac{d\bf r}{V} \, \Delta^\ast({\bf r})\Big( \frac{1}{|\lambda|} \nonumber \\
&-& \frac{T}{2}\sum_{\varepsilon_n,\sigma} \sum_{\bf p} \hat{K}_2(-i{\bf \nabla}_{\bf r}) \Big)\Delta({\bf r}), \nonumber\\
\hat{K}_2(-i{\bf \nabla}_{\bf r}) &=& |{\cal Y}({\bf p})|^2{\cal G}_{\varepsilon_n, \sigma}({\bf p}) \, {\cal G}_{-\varepsilon_n, -\sigma}(-{\bf p}-i{\bf \nabla}_{\bf r}),
\end{eqnarray}
where $\lambda$ denotes a coupling constant of a pairing interaction, ${\cal G}_{\varepsilon_n, \sigma}({\bf p})=(i\varepsilon_n - \xi({\bf p})+\sigma \mu H)^{-1}$ is the Gor'kov Green's function expressing a quasiparticle in the normal state, $\varepsilon_n$ is a fermionic Matsubara frequency, and $\nabla_{\bf r}$ denotes the gradient with respect to ${\bf r}$. By using the replacement $\sum_{\bf p} \rightarrow N(0)\int_{-\infty}^{\infty}d\xi({\bf p}) \langle \rangle_{\phi_p}$, we can carry out the summation over ${\bf p}$, and then, we find
\begin{eqnarray}
\sum_{\bf p}\hat{K}_2(-i{\bf \nabla}_{\bf r}) &=& N(0)\Big\langle \frac{2\pi i \, {\rm sgn}(\varepsilon_n)  |{\cal Y}({\bf p})|^2}{2i\,\varepsilon_n+2\sigma \mu H+i{\bf v}_F\cdot{\bf \nabla}_{\bf r}} \Big\rangle_{\phi_p} \nonumber\\
&=& N(0) \int_0^{\infty}d\rho \, 2\pi e^{-2|\varepsilon_n|\rho} \Big\langle |{\cal Y}({\bf p})|^2\nonumber\\
& \times& \exp\Big[ i\, {\rm sgn}(\varepsilon_n) \big(2 \sigma \mu H +i{\bf v}_F\cdot{\bf \nabla}_{\bf r}\big) \rho \Big] \Big\rangle_{\phi_p},
\end{eqnarray} 
where the equation $1/\alpha=\int_0^\infty d\rho \exp[-\alpha \rho] \quad ({\rm Re}[\alpha]>0)$ has been used. The summation over $\varepsilon_n$ and $\sigma$ yields 
\begin{eqnarray}
\frac{T}{2}\sum_{\varepsilon_n,\sigma,{\bf p}}\hat{K}_2(-i{\bf \nabla}_{\bf r}) &=& N(0)\int_0^\infty d\rho \, \frac{2\pi T \cos(2\mu H \rho)}{\sinh[2\pi T \rho]} \nonumber\\
&& \times \Big\langle |{\cal Y}({\bf p})|^2 \cos\big(-i{\bf v}_F\cdot{\bf \nabla}_{\bf r} \, \rho \big)\Big\rangle_{\phi_p}.
\end{eqnarray}
Since $T_{c0}$ is defined as the SC transition temperature at $D=\infty \quad (d^{-1}=0)$ and $H=0$, i.e.,
\begin{equation}
\frac{1}{|\lambda|}-N(0)T_{c0}\sum_{\varepsilon_n>0}^{\omega_c/T_{c0}}\frac{2\pi}{|\varepsilon_n|}=0,
\end{equation}
we have
\begin{equation}
\frac{1}{|\lambda|}\simeq N(0)\Big[ \ln \Big(\frac{T}{T_{c0}}\Big)+ \int_0^\infty d\rho \frac{2\pi T}{\sinh[2\pi T \rho]}\Big].
\end{equation}
Eventually, we obtain the GL quadratic term as
\begin{eqnarray}\label{eq:GL_2}
a&=& \sum_{\bf q}|C_{\bf q}|^2\bigg[\ln\Big(\frac{T}{T_{c0}}\Big)+\int_0^\infty d\rho \frac{2\pi T}{\sinh[2\pi T \rho]} \nonumber\\
&\times& \Big\langle |{\cal Y}({\bf p})|^2 \Big[1-\cos(2\mu H \rho ) \cos({\bf v}_F\cdot {\bf q} \, \rho)\Big]\Big\rangle_{\phi_p}\bigg],
\end{eqnarray}
where $C_{\bf q}$ is the coefficient of the Fourier expansion $\Delta({\bf r})=|\Delta(0)|\sum_{\bf q} \, C_{\bf q} \, \exp[i\, {\bf q}\cdot {\bf r}]$. For the film with the thickness $D$, the gap function takes the form of eq.(9), so that it follows that
\begin{equation}\label{eq:C_Q}
|\Delta(0)| C_{\bf q}=\Delta_{q_x} \, \big[ \delta_{q_y, \pi/D} + \delta_{q_y, -\pi/D}\big]/\sqrt{2} .
\end{equation}
Substituting eq.(\ref{eq:C_Q}) into eq.(\ref{eq:GL_2}), we obtain eq.(13). Note that the coefficient $a$ is given by the same expression for both of the FF and LO states.

\subsection{Expression of the GL quartic term}
The GL quartic term is given by 
\begin{widetext}
\begin{eqnarray}
b |\Delta(0)|^4&=&\frac{(2 \pi T_c)^2}{2 N(0)}\int \frac{d{\bf r}}{V} \, \hat{K}_4(\nabla_i) \Delta^\ast({\bf s}_1)\Delta({\bf s}_2) \Delta^\ast({\bf s}_3)\Delta({\bf s}_4) \big|_{{\bf s}_i\rightarrow{\bf r}}, \nonumber\\
\hat{K}_4(\nabla_i)&=& \frac{T}{2}\sum_{\varepsilon_n,\sigma} \sum_{\bf p} \, |{\cal Y}({\bf p})|^4 \, {\cal G}_{\varepsilon_n, \sigma}({\bf p}) \, {\cal G}_{-\varepsilon_n, -\sigma}(-{\bf p}+i\nabla_1)  {\cal G}_{-\varepsilon_n, -\sigma}(-{\bf p}-i\nabla_2) \, {\cal G}_{\varepsilon_n, \sigma}({\bf p}+i(\nabla_3+\nabla_2)), 
\end{eqnarray}  
\end{widetext}                
where $\nabla_i$ denoting the gradient with respect to ${\bf s}_i$ acts on $\Delta({\bf s}_i)$ \cite{Adachi}.
The summation over ${\bf p}$ can be performed in the same manner as that used in obtaining the quadratic term, and then, we obtain
\begin{eqnarray}\label{eq:quartic}
&&b= 2 \pi^2 T_c^2 \sum_{{\bf Q}_1+{\bf Q}_3={\bf Q}_2+{\bf Q}_4} C_{{\bf Q}_1}^\ast C_{{\bf Q}_2} C_{{\bf Q}_3}^\ast C_{{\bf Q}_4}  \nonumber\\
&&\times \prod_{j=1}^3\int_0^{\infty} d\rho_j \frac{2 \pi T \, \cos\big( 2 \mu H \, [\sum_{i=1}^3\rho_i ] \big)}{\sinh\big[2\pi T (\sum_{i=1}^3\rho_i)\big]} \Big\langle |{\cal Y}({\bf p})|^4 \nonumber\\
&& \times  \Big[  \cos\big( {\bf v}_F\cdot {\bf Q}_1 (\rho_1 + \rho_2) - {\bf v}_F\cdot {\bf Q}_2 \rho_2 + {\bf v}_F\cdot {\bf Q}_3(\rho_2+\rho_3) \big) \nonumber\\
&& + \cos\big( {\bf v}_F\cdot {\bf Q}_1 \rho_1 + {\bf v}_F\cdot {\bf Q}_2 \rho_2 + {\bf v}_F\cdot {\bf Q}_3 \rho_3 \big) \Big] \Big\rangle_{\phi_p},
\end{eqnarray} 
where ${\bf Q}_i=\big(Q_{x,i}, Q_{y,i}\big)$, ${\bf v}_F=2\pi T_c\xi_0 \big(\cos(\phi_p),\sin(\phi_p)\big)$. In eq. (\ref{eq:quartic}), the expression
\begin{equation}
C_{{\bf Q}_i} = \frac{\delta_{Q_{y,i}, \pi/D} + \delta_{Q_{y,i}, -\pi/D}}{\sqrt{2}}  \frac{\delta_{Q_{x,i}, q_x} + \delta_{Q_{x,i}, -q_x}}{\sqrt{2}} 
\end{equation}
is valid for the LO state with $\Delta({\bf r})= 2 \, |\Delta(0)|  \cos( y \pi/D)\cos( x q_x)$, while 
\begin{equation}
C_{{\bf Q}_i}=\frac{\delta_{Q_{y,i}, \pi/D} + \delta_{Q_{y,i}, -\pi/D}}{\sqrt{2}} \, \delta_{Q_{x,i}, q_x} 
\end{equation} 
should be used for the FF state in which $\Delta({\bf r})=\sqrt{2} \, |\Delta(0)|\cos(y \pi/D) e^{i \, x q_x }$.


\begin{thebibliography}{99}
\bibitem{com2} If the higher harmonics ${\rm cos}(n \pi y/D)$ ($n \geq 2$) is incorporated to describe the boundary condition (\ref{BC}), the appearance of this FF state in bulk limit seems to be suppressed.
\end{thebibliography}
\end{document}